\pdfoutput=1
\documentclass[aps, reprint, superscriptaddress, nofootinbib, showpacs,preprintnumbers,floatfix]{revtex4-1}
\usepackage{amsmath, latexsym, amssymb, hyperref, graphicx, color, multirow}
\usepackage[table]{xcolor}
\usepackage{tabularx}
\usepackage[T1]{fontenc}
\usepackage[capitalize]{cleveref}
\usepackage{booktabs} 
\usepackage{fouriernc} 
\definecolor{nicered}{rgb}{.7,.1,.1}
\definecolor{nicegreen}{rgb}{.1,.5,.1}
\definecolor{darkblue}{rgb}{0,0,.5}
\hypersetup{colorlinks, citecolor=nicegreen,linkcolor=nicered, urlcolor=darkblue}
 
\begin{document}
 
\title{Neutrino 2020: Theory Outlook $^{\S}$}

\author{Goran Senjanovi\'c  }

\affiliation{Arnold Sommerfeld Center, Ludwig-Maximilians University, Munich, Germany
} 
 
 \affiliation{ International Centre for Theoretical Physics, Trieste, Italy}

\begin{abstract}

I present a personal vision of what is essential in the field of neutrino mass, both from the point of view of what has been achieved and what could lie ahead. In the process, I offer a logical, theoretical and phenomenological rationale behind my opinions. It is however neither a summary of what was discussed in the conference nor a party-line viewpoint, rather an attempt to dig through the enormous body of material in our field in order to uncover a common unifying thread. The main focus is on the search for a predictive and self-contained theory of the origin and nature of neutrino mass, with the conclusion that the Left-Right Symmetric Model plays a special role in this aspect.

\end{abstract}

%\keywords{Left-Right symmetry; neutrino mass; Higgs mechanism.}

\maketitle

{
 \renewcommand{\thefootnote}%
   {\fnsymbol{footnote}}
 \footnotetext[4]{Based on the Theory Outlook talk, Neutrino 2020 International Conference on Neutrino Physics and Astrophysics, Fermilab, June 22-July 2, 2020}
}

%\subsection*{Disclaimer} \label{disclaimer}
%This is a rather personal vision of what is essential in the field of neutrino mass, both from the point of view of what has been achieved and what could lie ahead. It may appear biased, but in what follows I offer a logical, theoretical and phenomenological rationale behind my opinions. It is however neither a summary of what was discussed in the conference nor a party-line viewpoint - rather an attempt to dig through the enormous body of material in our field in order to uncover the common unifying thread.
   
\section{Prologue} \label{Prologue}

Neutrino 2020  will be remembered as historic. It was the biggest high-energy conference ever organized, with more than 3300 participants, and it was online due to the COVID-19 pandemic. In spite of the tragic circumstances, it achieved something special, unprecedented - it brought hundreds (at some moments more than 1000) of us together in real time, in the privacy and comfort of our homes, from all over the world.  It showed us that we may share great scientific excitement without polluting the environment, without having to leave our loved ones, or  interrupt our everyday life. \\

  If you were one of the attendees, you need no convincing that neutrino mass is special, for the fact that so many of us came together to discuss it speaks eloquently for itself. In any case, by being predicted to vanish in the Standard Model, neutrino mass is  arguably the best window into the new physics. And, as I discuss below, it touches into the core of it all.

\section{Neutrino mass: theory behind?} \label{neutrinomass}
 
 It took us a long time but we have come a long way. We are close to untangling neutrino masses and mixings - only the scale of the mass is missing, and the CP phase(s), as discussed amply during the conference. I have nothing meaningful to add to this profoundly important subject. I can only add my voice to the obvious: we must go on and complete the job in order to have an arena for testing the theory we are all eagerly awaiting for.\\
 
   We have a long way to go, though, since to get there demands that we come up with a self-contained, predictive theory of neutrino mass, testable and hopefully tested in near future. Anything else will imply falling short of this essential goal. And by a theory of neutrino mass, I do not mean tailor ordered models, but a theory that structurally predicts non-vanishing neutrino mass and relates it to new physical processes. An example is worth a thousand words, so let us go through the theories that actually did structurally predict neutrino mass before experiment. \\
 
 \subsection{ $SO(10)$ grand unified theory.}\label{so10}
 
 The $SO(10)$ grand unified model~\cite{Fritzsch:1974nn}
 is a minimal theory that unifies both the relevant particle interactions and a family of fermions. The building block, the 16-dimensional fermion representation, contains a right handed (RH) neutrino $N$ on top of the 15 up and down quarks, electron and neutrino. In the process of symmetry breaking $N$ becomes naturally heavy and equally naturally leads to small neutrino mass through the seesaw mechanism~\cite{Minkowski:1977sc}. This is a perfect example of a theory that predicted neutrino mass from its structure long before experiment, and can account for its smallness. The way it works is beautiful and worth recapitulating briefly here. The crucial point is that the unification of gauge couplings implies the existence of an intermediate scale below $10^{14} GeV$~\cite{Rizzo:1981jr} - the scale where $N$ gets the mass. This in turn guarantees that the naturally small neutrino mass is sufficiently large to explain atmospheric neutrino oscillations. \\

 The problem however comes from the main prediction of grand unified theories, besides the existence of magnetic monopoles, the proton decay. A successful theory should be able to predict proton decay branching ratios in a self-contained manner, without extra assumptions. Sounds nice in principle, but gets hard in practice. \\
 
   One is faced with the choice of whether to use small or large Higgs representations. Small  representations are clearly more appealing, but tend to lead to wrong fermion mass relations and typically require higher dimensional operators. This implies great many couplings and prevents making predictions regarding proton decay branching ratios. Larger representations can work at the renormalizable level, however the threshold effects become more pronounced which renders the predictions less certain.\\
   
 Moreover, large representations end up lowering the scale where gravity becomes important. What happens is the following. From black hole physics, one learns~\cite{Dvali:2007hz} that the the scale when gravity gets to be strong  (call it $\Lambda_{strong}$) is not $M_{Pl}$ as we would normally expect, but depends on the number of degrees of freedom - or number of species $N_{species}$ as coined by Dvali ~\cite{Dvali:2007hz} - of the theory
 \begin{equation}
 \Lambda_{strong} = \frac{M_{Pl}}{\sqrt{N_{species}}}\,.
 \end{equation}
%Moreover, it is argued in~\cite{Dvali:2007hz} that  this cannot be be eliminated by resummation, since it is a pure non-perturbative effect. 
\\

 In $SO(10)$,  $N_{species} \sim 10^2-10^3$, significantly lowering the scale. In order for the GUT theory to be predictive, however,  $\Lambda_{strong}$ should be much bigger than the unification scale $M_{\rm GUT} \simeq10^{16}$GeV, 
 making it even more problematic and at the same time less convincing to ignore the higher dimensional operators. And maybe even worse, large representations can lead~\cite{gg}  to collective processes that could violate the unitarity, forcing a change of the regime of the theory~\cite{Dvali:2020wqi}.\\
 
    Bottom line: as beautiful and profound the $SO(10)$ grand unified theory is, no clear cut predictions have been obtained that could lead to a smoking gun signature. No universally accepted model has emerged yet and in its absence we go on with our list of potentially self-contained, predictive theories.

  \subsection{Supersymmetric Standard Model}\label{SSM}
  
  Another clear example of a theory that predicts neutrino mass from its structure is the Supersymmetric extension of the Standard Model (SSM). The role of the RH neutrino is played here by the photino and Zino, the partners of the photon and the Z-boson, and the seesaw mechanism is realised through the small vev of the sneutrino, the supersymmetric scalar partner of the neutrino. This amounts to not imposing the so-called R-parity, a symmetry assigning opposite sign on particles and their super partners. Since there is nothing natural {\it a priori} about imposing such a symmetry by hand, neutrino mass is a natural generic feature of the low energy supersymmetry.\\
  
    This sounds nice in principle, but in practice it is a nightmare, as with any other phenomenological issue within the SSM - there are far too many parameters to make any coherent statement. In reality the SSM is more like an endless collection of different models rather than  a well defined theory. So one normally fixes the parameter space in order to make 'predictions', the very  opposite of the way a theory should work. \\
  
  I say this with a heavy heart, since I have a vested interest in low energy supersymmetry, regarding the prediction of gauge coupling unification. Following the work of Dimopoulos et.al. \cite{Dimopoulos:1981yj}, Marciano and I \cite{Marciano:1981un} showed that unification requires a heavy top quark with a mass around 200 GeV. Moreover, the unification of gauge couplings was predicted ten years before LEP, at the time when the weak mixing angle was wrongly measured - supersymmetry actually anticipated experiment by a long shot.\\
  
  It is well known that the SSM could have a natural dark matter candidate if the lightest neutralino was to be stable, or in other words, if R-parity was somehow magically conserved to an astonishing precision. This can be of course postulated by hand - and is often done - but it can be actually automatically achieved in a large class of theories, with B-L gauge symmetry spontaneously broken at some large scale by a field(s) with even charge~\cite{Aulakh:1999cd}. A prime example is provided by a supersymmetric $SO(10)$ theory that employs a $126_H$ Higgs field for this purpose~\cite{Aulakh:2000sn}. \\
  
  The argument is somewhat subtle and goes as follows. The first thing to notice is that in the SSM, R-parity cannot be spontaneously broken since it would lead to the existence of a Goldstone boson, the Majoron, coupled to the Z boson \cite{Aulakh:1982yn} - forbidden by the Z width. The point is that imposing R-parity implies a continuous global B-L symmetry. Secondly, if a theory with a new high scale possesses an automatic gauged B-L symmetry, R-parity (automatic consequence of B-L) is either broken at the large scale or not. In the former case one must ensure that the breaking in the low energy sector is small enough to be compatible with experiment, and here we have nothing to say on this.\\
  
   The latter case is more profound since R-parity either remains unbroken or gets broken spontaneously at low energies through a vev of LH sneutrino. However, since by assumption the new physics scale beyond SSM is large, this implies that the corrections to the Majoron mass must be small, suppressed by a low supersymmetry breaking scale - and thus a theory with a large scale would lead un unacceptable light pseudo-Goldstone~\cite{Aulakh:1999cd}. In other words, in a theory with automatic gauged R-parity, unless R-parity is broken at the high scale, it is guaranteed to remain exact, implying a stable dark matter candidate. A  physical realization of this appealing picture is a renormalisable seesaw mechanism where the RH neutrino (and sneutrino) gets a large mass through spontaneous symmetry breaking of B-L symmetry~\cite{Aulakh:1999cd}. A natural arena behind it is provided by the Left-Right symmetry~\cite{Aulakh:1998nn}, besides the above mentioned $SO(10)$ grand unified theory.\\
   
    In short, low energy supersymmetry has surely great merits, however, it fails as a self-contained theory of neutrino mass. Since low energy supersymmetry is tailor made for grand unification, one could hope that the supersymmetric GUT models could do the job. The problem again is proton decay, and the situation is much worse that in the non-supersymmetric case, since there is a new important contribution from super-partners, that tends to dominate the proton decay rate. And just as in the ordinary grand unification, no universally accepted model emerged, and no precise predictions.
    We are forced to move on.

 \subsection{ Left-Right Symmetric Theory}\label{LR}
 
   So we turn our attention to a much more promising candidate for a theory of neutrino mass, the Left-Right Symmetric Model (LRSM). It was proposed originally~\cite{Pati:1974yy} to account for parity violation in weak interactions by attributing it to the spontaneous symmetry breaking of an otherwise perfectly symmetric world - hence the name. This brings Left-Right symmetry breaking on the same footing of gauge symmetries, which is appealing {\it per se}. But more important, since if there is left, there must be right, and the theory predicts from the onset a right handed neutrino - and 
    thus also neutrino mass. \\
    
    Neutrino mass seemed a curse when the theory was proposed in the '70s, since neutrino still seemed massless and the SM wanted it massless. It was moreover hard to understand why it was so light compared to the electron. The seesaw mechanism turned the curse into a blessing as the solar neutrino puzzle started, slowly but surely, to call for neutrino mass. It not only made the neutrino mass naturally small, but  connected its smallness to the near maximality of parity violation in the weak interaction. \\
    
    The crucial aspect lies in the word 'near' - everything works out great as long as you don't take the scale of parity restoration to infinity as in the SM. It took some forty years to show it, but finally it emerged that the LRSM in its minimal version  is actually a self-contained, predictive theory of both the origin and the nature of 
  neutrino mass. \\

   We shall returns to this strong claim below and make sure we justify it. But before we go on, an important question must be raised as to whether or not we can ignore gravity in all this.
  
   \subsection{ Does gravity matter?}\label{gravity}
    
 We all know that matter gravitates, but we also believe that gravity does not matter at today's energies. Actually, some years ago when solar neutrinos were thought to oscillate on their way towards us - the so-called 'just so' oscillations - it was suggested that  Planck-scale-suppressed effects could actually account for a tiny neutrino mass~\cite{Akhmedov:1992hh}. We know today, however, that neutrino mass differences are too large for Planck scale suppressed physics to matter, and it would appear that gravity can be safely ignored.\\
 
   However, the situation turns out to be more subtle. It has been argued~\cite{Dvali:2013cpa}, in analogy with QCD, that the gravitational anomaly can induce chiral symmetry breaking with neutrino taking the role of up and down quarks in QCD
  \begin{equation}\label{nucondensate}
\langle \bar \nu \nu \rangle = \Lambda_{gravity}^3\,.
\end{equation}
The relevant scale $\Lambda_{gravity}$ depends only on the Planck scale $M_P$ and is exponentially suppressed by the number of degrees of freedom (number of species) involved in the theory~\cite{Dvali:2017mpy}
\begin{equation}\label{nuscale}
\Lambda_{gravity}^3 \lesssim M_{Pl}^3 \, e^{-N_{species}}\,,
\end{equation}
 where $N_{species}$ here counts the real degrees of freedom. In the SM, summing up the three generations of fermions, the strong and electro-weak gauge bosons, and the Higgs doublet, gives $N_{species}^{SM} = 118$. Adding the three RH neutrinos assumed above would give $N_{species}=124$ and $\Lambda_{gravity} \lesssim GeV$, which can surely impact neutrino mass. In principle it could also affect the electron mass and it should be kept in mind as the SM Higgs mechanism keeps being verified for lighter generations.

\section{The core of it all} \label{core}

In his classic work~\cite{Majorana:1937vz}, Majorana left us a lasting legacy by showing that a neutral fermion can be its own antiparticle through the so-called Majorana mass term. The only candidate for a Majorana fermion in the SM is the neutrino, and  if it is so, it would imply Lepton Number Violation (LNV) by two units. There are two fundamental $\Delta L = 2$  processes, at low and high energies, respectively:

\begin{itemize}
\item neutrinoless double beta decay

\item LNV at hadron colliders (LHC)
\end{itemize}

It is impossible to overemphasise the importance of these processes, and in what follows we shall focus a lot on the profound connection between them. We will show how the observation of the former could actually imply observable effects of the latter at the LHC or a future hadron collider. This is arguably our best door to  new physics at accessible energies. 

 \subsection{ Neutrinoless double beta decay}\label{0nu2beta}
 
\begin{figure}[h]
\centerline{
\includegraphics[width=.5\columnwidth]{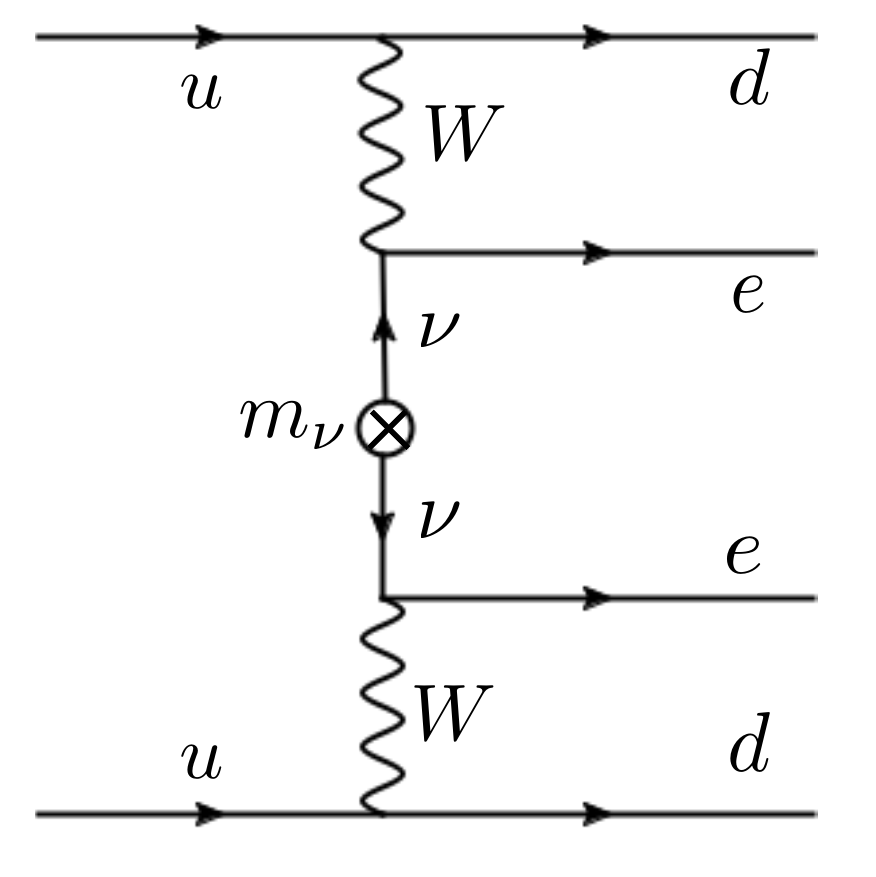}  }
\caption{Neutrinoless double-beta decay through the exchange of Majorana neutrino.}
\label{fig-0nu2beta}
\end{figure}

 This text-book example~\cite{Racah:1937qq} of low energy LNV has become the central experimental and theoretical focal point in our search for the nature of neutrino mass. If neutrino mass was of Majorana nature, it would allow for the process shown in Fig. \ref{fig-0nu2beta}.
 It is easy to see that the amplitude for such process is given by 
  \begin{equation}\label{Anu}
{\cal A}_{\nu} \propto G_F^2 \frac{m_\nu^{e e}}{p^2}\,,
 \end{equation}
 where $m_\nu^{e e}$ is the appropriate neutrino mass matrix element, and $p \simeq 100 MeV$ is a measure of neutrino virtuality, relevant for the nuclei in question. From the non observation of $0 \nu 2 \beta$, 
 or more precisely from the limit $\tau_{0 \nu 2 \beta} \gtrsim 10^{26} \, yr$, set recently by the GERDA experiment~\cite{Agostini:2020xta}, one obtains an upper limit on the neutrino Majorana mass, $m_\nu^{e e} \lesssim 0.1 \, eV$. This is  exciting since it is getting close to the scale of neutrino mass from atmospheric oscillations. \\
 
Keep in mind that the neutrino Majorana mass contribution in Fig. \ref{fig-0nu2beta} goes through the usual weak interaction, and thus both electrons come out left-handed. \\

 It is often said that this process is a probe of neutrino Majorana mass, that it would surely hint that neutrino could be Majorana if observed. There is even a claim (known as the black-box theorem~\cite{Schechter:1981bd}) that neutrino would {\it have} to be Majorana.
This would be very exciting if it was not wrong. First of all,  although this process indicates the Majorana possibility, the actual contribution from the so-called $0 \nu 2 \beta$ black box is negligibly small, giving 
$m_\nu \simeq 10^{-24}\, eV$~\cite{Duerr:2011zd}, zero for all practical purposes. That  neutrinoless double beta decay may not be due to neutrino Majorana mass, but rather to some new unknown physics, was pointed out originally more than sixty years ago~\cite{maurice}, and yet, it is often ignored to this day. \\

 In order to appreciate the impact of new physics, arriving from a new large scale $\Lambda_{\not L}$, let us write a relevant effective operator describing the $0 \nu 2 \beta$. Being a six-fermion operator, it has dimension nine
  \begin{equation}\label{d=9}
{\cal A}_{NP} \propto \frac {1} {\Lambda_{\not L}^5} (e \,e \,u \,u \, \bar d\, \bar d )\,.
 \end{equation}
Compare it with the neutrino contribution in \eqref{Anu}, which as we said probes the value $m_\nu \simeq 10^{-1} eV$. 
 It implies in turn $\Lambda_{\not L} \gtrsim  3TeV$ - tailor-made for the LHC.  Recall that a similar limit on the scale of baryon number violation is $\Lambda_{\not B} \gtrsim  10^{16} GeV$, astronomically far from direct experimental reach.   \\

  Moreover, there are clear ways of knowing that  new physics and not neutrino mass is the source of the process: it is sufficient that one of the electrons that comes out of $0 \nu 2 \beta$ is right-handed. In this case, $0 \nu 2 \beta$ could actually be a probe of the theory of neutrino mass, rather than of the mass itself - a far more exciting perspective.\\
   
  In summary, if new physics were to cause neutrinoless double decay, it could lie tantalisingly close to the LHC energies. Hard to imagine a better motivation for observable new physics, suggested by pure phenomenological considerations - and yet it is rarely mentioned. This possibility would be even more exciting for it would open the door to lepton number violating processes at today's or near future hadron colliders.

 \subsection{LNV violation at colliders: The Keung-Senjanovi\'c (KS) process}     The new physics that we have been talking about  is direct lepton number violation at hadron colliders, in the form of same sign lepton pairs accompanied by jets, as argued by Keung and myself~\cite{Keung:1983uu} more than thirty five years ago. This is a high energy analog of the neutrinoless double beta decay, with the same lepton number violation pattern by two units. It also probes lepton flavour violation, and has an advantage over the neutrinoless double beta decay: Unlike the low energy situation, in this case one can probe directly the physics behind it. The prime candidate is the Left-Right symmetric theory, but not exclusively: Most theories of neutrino Majorana mass lead to the KS process, which makes it a paradigm for lepton number violation at hadron colliders.
 
 It took a long time for this process be universally recognised, and the LHC has finally made it a reality - today both ATLAS~\cite{Aaboud:2019wfg}  and CMS~\cite{Sirunyan:2018pom} are actively pursuing it. There is a profound connection between the KS process and neutrinoless double beta decay, that we discuss below. But to see what it is all about, we first need to step back.

  \section{Back to basics: parity and the SM} \label{basics}  
      
  In  a sense, it all started with the bombshell of Lee and Yang in 1956 that parity could be breaking in weak interactions~\cite{Lee:1956qn}. When experiment some six month later confirmed it, and moreover, showed that the breaking is maximal, the road to the SM was paved. It led to the $V - A$ theory~\cite{Sudarshan:1958vf}, with a pure left-handed weak current and the total absence of Left-Right symmetry
   \begin{equation}
J_\mu^W = \overline u_L \gamma_\mu d_L + \overline \nu_L \gamma_\mu e_L\,.
   \end{equation}\\

  On the other hand, Left-Right symmetry  is a fundamental space-time symmetry and it shapes our understanding of the world. Its  role is similar to the one played by Lorentz symmetry or even more basically by the invariance under the Galilean laws of mechanics. It is perhaps not surprising that even Lee and Yang in their classic seemed reluctant to accept that Left-Right symmetry could be broken at the fundamental level. In their own words~\cite{Lee:1956qn}:  ``If such asymmetry is indeed found, the question could still be raised whether there could not exist corresponding elementary particles exhibiting opposite asymmetry such that in the broader sense there will still be over-all right-left symmetry". \\
       
   In the SM, maximal Parity violation is essential, in a way as important as the fact that the weak current has vector  and axial vector character. The latter allows for a gauge theory, the central step taken by Glashow, the one that made all the difference. But is is the former one that leads to the simple and predictive Higgs mechanism origin of mass. 
   Maximal parity violation  implies LH fermion doublets and RH singlets,
   leading to massless quarks and charged leptons. This is cured with  the single Higgs doublet for both leptons and
   quarks.  \\
   
   The SM, which started as a theory of weak interactions, turned into a theory of the origin of mass. Mass becomes a dynamical variable whose value determines uniquely the associated Higgs boson decay. Today we know with certainty that $W$, $Z$ and the third generation charged fermions get the mass from the Higgs mechanism, and even the muon is getting there~\cite{Sirunyan:2020two}. \\
   
     In order to truly appreciate the importance of the maximal parity violation for the sake of the SM success, imagine for a moment that Lee and Yang had been wrong and that we live in a parity conserving world with the LR weak interaction, or in other words, the vector-like weak currents. All hell would break loose as we shall see\footnote{I advise students (and even  teachers) of the SM to go through this exercise - it is mind boggling how often the SM gets criticised in spite of its remarkable simplicity and predictivity.}.\\
     
       In such imaginary vector-like version of the SM, based on the $SU(2) \times U(1)$ gauge symmetry, the quark assignment then becomes
    \begin{eqnarray}
q_L = \left( \begin{array}{c} u_L \\ d_L \end{array}\right), \,\,\,\,\,\,\,\,
%& \stackrel{P}{\longleftrightarrow}&\,\,\,\,\,
 \left( \begin{array}{c} u_R \\ d_R \end{array}\right)= q_R,
 \label{ds21}
\end{eqnarray}
which clearly allows for a direct quark mass term $\overline q_L M q_R$. \\
   
   This is its doom. The first issue is the  miracle that $M$, a gauge singlet, is of the order of $M_W$, the scale of symmetry breaking, instead of escaping to a large scale as say $M_P$. Well, miracles happen, you say, but the 
   problem is that  up quark and down quarks mass matrices would be the same. The only way out would be the inclusion of an adjoint Higgs, a real triplet $T$, with the following Yukawa sector
   \begin{equation}
   {\cal L}_Y = \overline{q_L} ( M + Y_T T ) q_R + h.c.\\
   \end{equation}

 Once the triplet gets a non-vanishing vev  $\langle T \rangle = v  \rm{diag} (1,-1)$, the up and down quark mass matrices get split
   \begin{equation}
M_u = M + Y_T v, \,\,\,\,\,\, M_u = M - Y_T v\,,
   \end{equation}
and seemingly all is well. However, flavor conservation in neutral currents requires another little miracle, the alignment of $M$ and $Y_T$ matrices, since they must be simultaneously diagonalised. Well, miracles happen, you say again, but the problem then becomes that $M_u$ and $M_d$ get simultaneously diagonalised as well, implying the vanishing of the quark mixings, $V_{\rm CKM} = 1$.  Notice that you would get nothing by adding yet another triplet $T'$, for you can always rotate them so that just one has the vev - and only real triplets can have the Yukawas. In this theory there would be no quark mixing, not sufficiently to describe the real world.\\
 
 It gets even worse. The real triplet Higgs vev would give $M_Z=0$, requiring an additional, different Higgs multiplet (say a doublet) on top - basically, all the predictions of the SM would be gone. In this sense the great success of the SM appears even more profound. And it was maximal parity violation, as we keep stressing, that made all the difference. Weinberg put it nicely some ten years ago: ``V-A was the key"~\cite{Weinberg:2009zz} to make the SM what it is. The maximality of parity violation is essential for the success of the SM, both in the gauge sector and regarding charged fermion masses.\\
 
 However, the same maximal parity violation, together with  minimality, forbids neutrino mass by the very SM structure in the lepton sector
    \begin{eqnarray}
 \left( \begin{array}{c} \nu_L \\ e_L \end{array}\right)\,,  \,\,\,\,\,\,\,\,\,\,
%& \stackrel{P}{\longleftrightarrow}&
 e_R\,.
 \label{nueSM}
\end{eqnarray}
It must be stressed that  minimality plays here an essential role. It excludes the right-handed neutrino, and the Higgs doublet is insufficient to give neutrino an otherwise Majorana mass. And minimality in true theories is a must: without it, the theory would be stripped of predictivity. The vanishing of neutrino mass is a true, profound prediction of the SM.
With the discovery of neutrino mass, the SM finally showed its incompleteness, one that paves the way to new physics.\\

 In this sense, our imagined Left-Right symmetry, that failed miserably in the quark and gauge boson sector, succeeds in the leptonic sector. The vector-like lepton world would look like  
 \begin{eqnarray}
 \left( \begin{array}{c} \nu_L \\ e_L \end{array}\right)\,,  \,\,\,\,\,\,\,\,\,\,
%& \stackrel{P}{\longleftrightarrow}&
 \left( \begin{array}{c} \nu_R \\ e_R \end{array}\right),
 \label{nue21}
\end{eqnarray}
so that the RH neutrino would be automatic, and neutrino mass  ensured. \\

 Let me pause to reflect on what this taught us. We ended up with a kind of conflict, with neutrino mass calling for Left-Right symmetry and charged fermions demanding it to be broken maximally. 
The way out of this conflict, as always when you need a symmetry and need it broken at the same time, is to have the symmetry {\it spontaneously} broken.  This turns out to make all the difference, and allows for the development of a true theory of neutral fermion masses, the Left-Right Symmetric Model. From the modern point of view, it appears almost inevitable, although the historic route was rather different, basically opposite. The Left-Right symmetric theory was originally suggested in order to account for parity violation in weak interaction, nothing more. As we said, the prediction of non-zero neutrino mass seemed to be its curse, until history vindicated it and turned it into its main virtue.

 \section{The theory} \label{theory}
 
The minimal LRSM ~\cite{Pati:1974yy} is based on the gauge group 
$G_{LR}= SU(2)_L \times SU(2)_R  \times U(1)_{B-L}$.
Thus a new set of gauge bosons, the right-handed ones, are added to $W$ and $Z$, which now are called the left-handed ones.
Quarks and leptons are completely symmetric under parity:
\begin{eqnarray}
Q_L = \left( \begin{array}{c} u \\ d \end{array}\right)_L
& \stackrel{P}{\longleftrightarrow}&
 \left( \begin{array}{c} u \\ d \end{array}\right)_R = Q_R\,,
 \nonumber \\
\ell_L = \left( \begin{array}{c} \nu \\ e \end{array}\right)_L
& \stackrel{P}{\longleftrightarrow}&
 \left( \begin{array}{c} \nu \\ e \end{array}\right)_R = \ell_R\,.
\label{dsLR21}
\end{eqnarray}
\\

Strictly speaking, one could use charge conjugation instead of parity and  a lover of grand unification may prefer it, since in the SO(10) grand unified theory it is actually a finite gauge 
 transformation~\cite{Slansky:1981yr}. 
 She should rest assured that most of the results go through with proper  modifications. In what follows we will be however mostly focused on parity, unless specified otherwise - after all, it is parity that first come to mind when one speaks of left transforming into right.

    LR symmetry delivers a fundamental prediction of neutrino mass from the outset, just as it would have been in an imagined world of  vector-like SM: The existence of $\nu_L$ implies by default the existence of $\nu_R$. The issue of the smallness of neutrino mass disfavours the Dirac case~\cite{Branco:1978bz}, and points to the Majorana version through the seesaw mechanism, a natural outcome of symmetry breaking as we now describe. \\

   The following step is crucial.  Parity gets broken spontaneously, and the scale of breaking corresponds to the mass of the $SU(2)_R$ gauge bosons $W_R$, with $M_{W_R} \gg M_{W_L} $ (the SM is recovered for $M_{W_R} \to \infty $).  
%   We refer the reader to~\cite{Senjanovic:2011zz} for details on the Higgs sector and symmetry breaking in this model. 
Most important, in the process of symmetry breaking, the RH neutrino (called $N$ hereafter) becomes heavy,   $M_N \propto M_{W_R}$. It becomes a Majorana neutral lepton, with its own dynamics. 
%It is no longer an almost  phantom particle, a singlet added to the SM just in order to give neutrino a mass, whose only interaction besides gravity is with the LH neutrino through the Dirac Yukawa coupling. 
In the LRSM, $N$ has interactions with $W_R$, and thus can be naturally produced in a hadron collider - it leads to testable predictions of new phenomena.\\

And most important, in the process of symmetry breaking one predicts $M_N \propto M_{W_R}$, making the RH neutrino $N$ naturally heavy. The seesaw formula
 \begin{equation}\label{seesaw}
M_\nu = - M_D^T \frac {1}{M_N} M_D
 \end{equation}
then ensures the lightness of neutrino, connecting it to the near maximality of parity violation in nature. An important comment: one often talks of producing $N$ at colliders through the mixing with light neutrino {\it via} $M_D$, but that is not the seesaw picture where $M_N \gg M_D$. Rather, it would require cancellations and the accidental smallness of neutrino mass.\\

There is more to it, though. It turns out that the theory is self-contained, in the sense that the seesaw can be untangled and $M_D$ determined~\cite{Nemevsek:2012iq} from \eqref{seesaw}. This is best illustrated for the case of real $M_D$ (or alternatively, the use of charge conjugation instead of parity), which then gives
 \begin{equation}\label{untangle}
 M_D = i M_N \sqrt  {M_N^{-1} M_\nu}\,.
 \end{equation}
All ambiguity that normally plagues~\cite{Casas:2001sr} the seesaw is gone.\\

The key role in this is played by the KS process that gives LNV at colliders, as advertised above, and at the same time, provides a direct test of the Majorana nature of $N$.  It moreover allows to determine $M_N$, which together from $M_\nu$, suffices to predict $M_D$ from \eqref{untangle} - and in turn the associated $N$ decays, discussed below.
  
  \subsection{The KS process: from Majorana to LHC}   
  
\begin{figure}[h]
\centerline{
\includegraphics[width=.7\columnwidth]{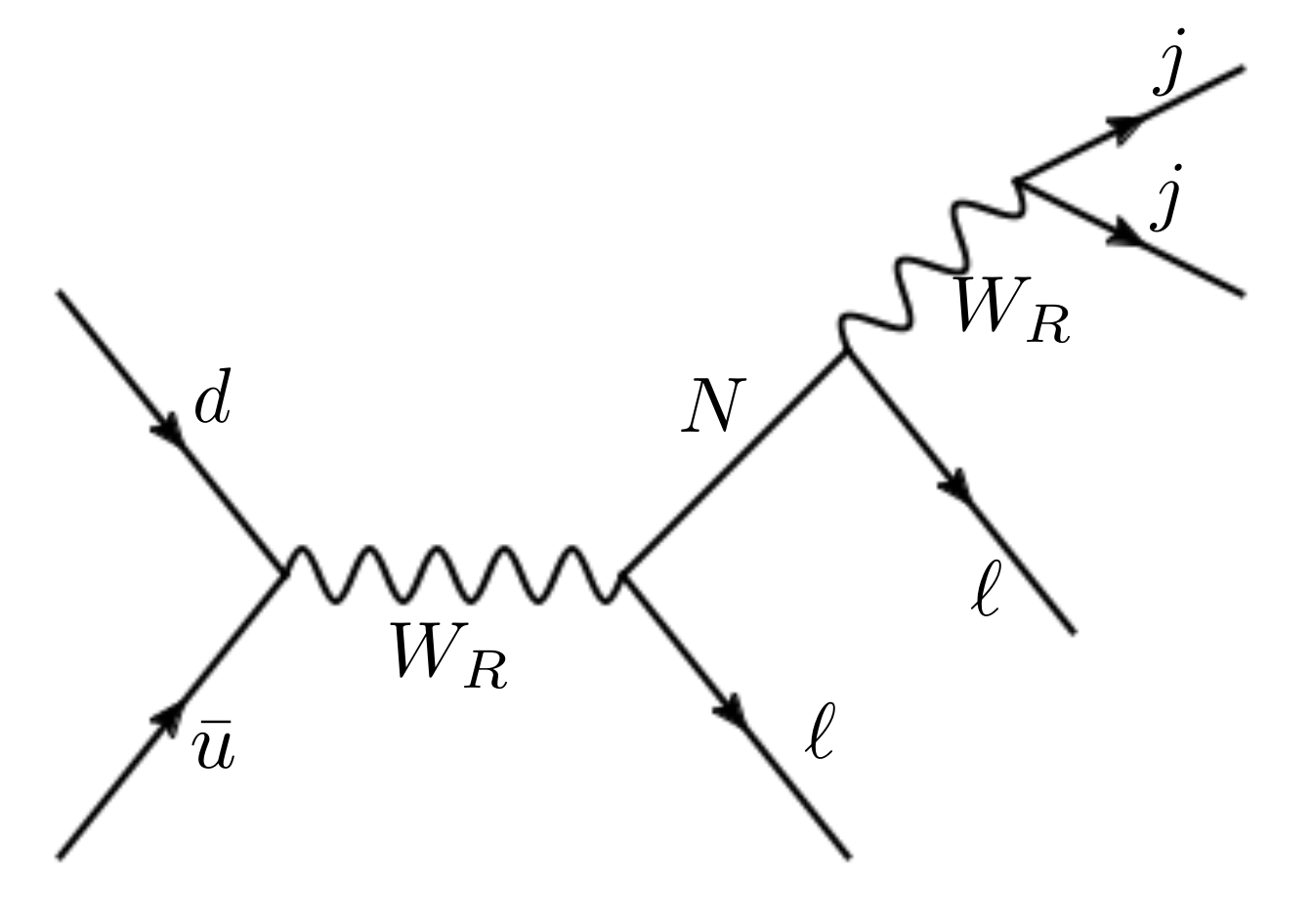}  }
\caption{Keung-Senjanovi\'c process: LNV at colliders, consisting of two same sign leptons and two jets.}
\label{fig-ks}
\end{figure}

The KS process has two important aspects. The first one, the direct LNV at hadron colliders is shown in  Fig.\ref{fig-ks}.
%
 
%\begin{figure}[h]
%\centerline{
%\includegraphics[width=.7\columnwidth]{ks2-diagram.pdf}  }
%\caption{KS process, showing two possible outcomes:  two same sign leptons or  two opposite sign charged leptons.}
%\label{fig-ks}
%\end{figure}
%
In complete analogy with the (LH) $W$ production, once $W_R$ is produced it decays into a lepton and a RH neutrino $N$. Because $N$ is a Majorana particle, this can lead to LNV, just as in $0\nu2\beta$, and you again get same-sign leptons. We should stress that this feature is rather generic of models of neutrino Majorana mass, for a review see~\cite{Senjanovic:2011zz}. This said, one can always imagine situations where the Majorana mass spectrum conspires to create Dirac-like states~\cite{Gluza:2016qqv} - after all, a Dirac particle is a case of two degenerate Majorana masses. Not being generic, we shall discuss such particular situations in what follows.  \\

However, in contrast with LNV in $0\nu2\beta$, here we have a collider process, where by measuring energies and momenta of the final state particles, one can reconstruct the properties of $W_R$ and $N$.  The KS  process allows to probe $M_N$, complementing the low energy determination of $M_\nu$. From $M_D$ and $M_N$ one in turns obtains the $\nu - N$ mixing, which then allows to determine the relevant associated decay rates.\\

 We  also have a deep connection with $0\nu 2\beta$~\cite{Tello:2010am}, which makes $0\nu 2\beta$ crucial for the probe of the theory. If at least one electron comes out RH, it would imply that in the LRSM it is the physics of $W_R$ and $N$ that dominates this process and would provide a great impetus for its accessibility at the LHC or the next hadron collider. For a recent study regarding the neutrinoless double beta decay in the LRSM, see~\cite{Li:2020flq}.

There is moreover a connection with low energy lepton flavor violation processes, such as $\mu \to e \gamma$, $\mu \to e \,e \, \bar e$ and $\mu \to e $ conversion in nuclei.  These rare processes are actually correlated in the LRSM~\cite{Cirigliano:2004mv}. For the possibly accessible LR symmetric scale in the TeV region, their suppression implies $m_N \ll M_{W_R}$ (or an unnatural degeneracy of $N$ masses), in analogy with the small charm quark mass as a guarantee for tiny $K - \bar K$ mass difference. This has been nicely summarized in the Ph.D. Thesis by Tello~\cite{Tello:2012qda}.\\

\begin{figure}[h]
\centerline{
\includegraphics[width=.7\columnwidth]{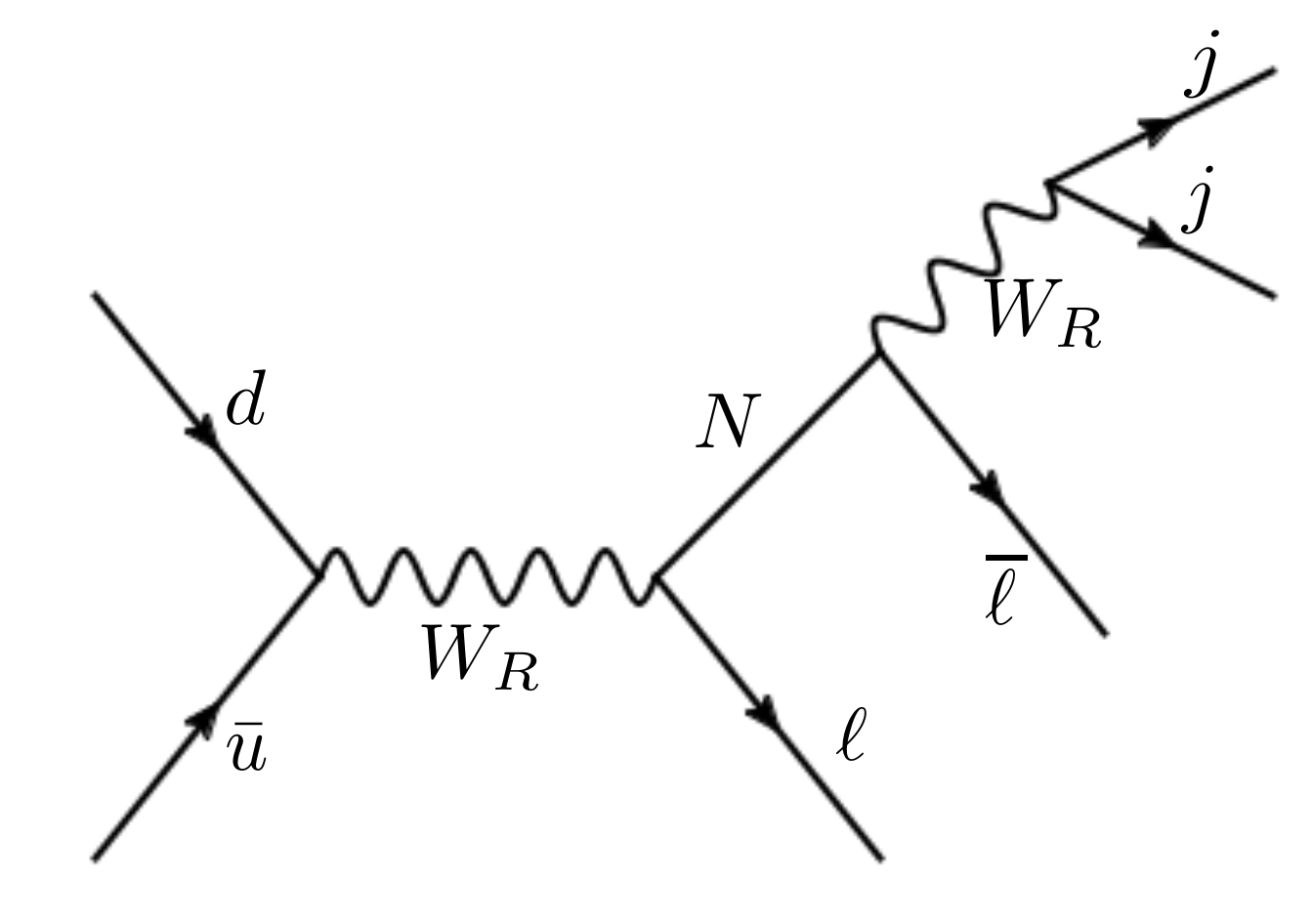}  }
\caption{Keung-Senjanovi\'c process: Lepton number conserving version, with two opposite sign charged leptons and two jets.}
\label{fig-antiks-2}
\end{figure}

There is still more to it. Since $N$ is Majorana, half particle and half anti-particle, once you produce it on-shell it has to decay equally often into a charged lepton or its anti-particle~\cite{Keung:1983uu}. The outcome of the process is 50\% same-sign leptons and 50\% lepton-antilepton, as depicted in Fig.
\ref{fig-antiks-2}.
%\ref{fig-ks} 
This offers a unique opportunity of probing the nature of mass: we can literally see inside Majorana's vision. \\

Before moving on to discuss the challenge of disentangling the seesaw experimentally, a comment is called regarding the production of $N$ in the KS process. Clearly, the existence of $W_R$ is tailor made for this task, but is it crucial?  One could argue that $W_L$ could do the job - after all, $N$ couples to it through the mixing $\theta_{\nu N} = M_N^{-1} M_D$. In the LRSM this is however negligible since the production rate goes as $\theta_{\nu N}^2 = M_\nu /M_N$ from \eqref{untangle}. Once again, the predictivity of the theory helps keeping things clear and simple. 

What would happen, however, if one were to disregard the theory behind its existence and just kept $N$?
Due to the ambiguity~\cite{Casas:2001sr} of the seesaw in the SM, one could now artificially boost arbitrary $M_D$ to make the production as large as needed~\cite{Datta:1993nm}. In other words, one could profit from the failure of the seesaw to make predictions {\it per se}. 

The trouble is that this then goes against the whole point of providing a rationale behind a small neutrino mass, since neutrino would be light due to accidental cancellations, rather than a small $\theta_{\nu N}$ mixing. The LR symmetry and the existence of $W_R$  play a fundamental role in making a seesaw scenario a true theory of neutrino mass.

 \subsection{LR symmetry: untangling the seesaw}  

Once the Majorana nature of $N$ is verified,  one can turn to probing the nature of the LH neutrino mass, by disentangling the see-saw.  This has been studied in great detail in recent years~\cite{Senjanovic:2018xtu}. By knowing the masses of RH and LH neutrinos, one can determine the neutrino Dirac mass matrix $M_D$ as illustrated in \eqref{untangle}, and then in turn predict a plethora of new decays. In particular, one can predict the decay of RH neutrino into a $W$ and a charged lepton, illustrated here for a simple case of the same LH and RH leptonic mixings
\begin{equation} \label{Ndecays}
\Gamma(N \to W \ell) \propto \frac{m_N^2}{M_W^2} \, m_\nu\,.
\end{equation}
For an associated LHC study, see e.g.~\cite{Arbelaez:2017zqq}.\\

 Since this decay is induced by the $\nu - N$ mixing, the outgoing charged lepton must be left-handed. There is also an associated decay into the right-handed charged lepton due to $W_L - W_R$ mixing, and so it is important to be able to measure the lepton chirality. This was studied in~\cite{Ferrari:2000sp}, and argued to be feasible at the LHC.\\

Notice the key point here: this is in complete analogy with the SM. In the SM, the knowledge of charged fermion masses gives you the Higgs decay rates
\begin{equation} \label{hdecays}
\Gamma(h \to \bar f f) \propto \frac{m_f^2}{M_W^2} \, m_h\,.
\end{equation}
So one can say that LRSM does for uncharged fermions what the SM did for charged fermions - allows to probe the Higgs origin of mass. \\

It is hard to overemphasise the importance of this result - it means that the theory is self-contained. Compare this with attempts  to use additional ad-hoc textures for the same task. To appreciate what it means, imagine giving up the minimality of the SM and its prediction of Yukawa couplings - and thus the Higgs decay rates - from the knowledge of quark and charged lepton masses. Instead, you decide to add more Higgs and look for the textures that would give you predictions, of the same type that you had in the first place and you gave up.\\

Alternatively, one sometimes makes assumptions about the parameter space of the theory, and then makes custom-fit adjustments by enlarging the minimal model - and sadly, loses the main predictions that enable to disentangle the seesaw. This is in a sense even more problematic than the texture tricks, since the last thing to do is fix the parameters and then look ways to live with them, especially when you have at hands a real theory that  gives you a handle on the parameter space in question. So again, imagine giving up the minimality of the SM that determines the charm quark mass from $K - \bar K$ mass difference, and instead fixing the mass arbitrarily and looking for extensions that can account for it. In a sense, this should be considered opposite from predictivity. \\

\begin{figure}[h]
\centerline{
\includegraphics[width=.7\columnwidth]{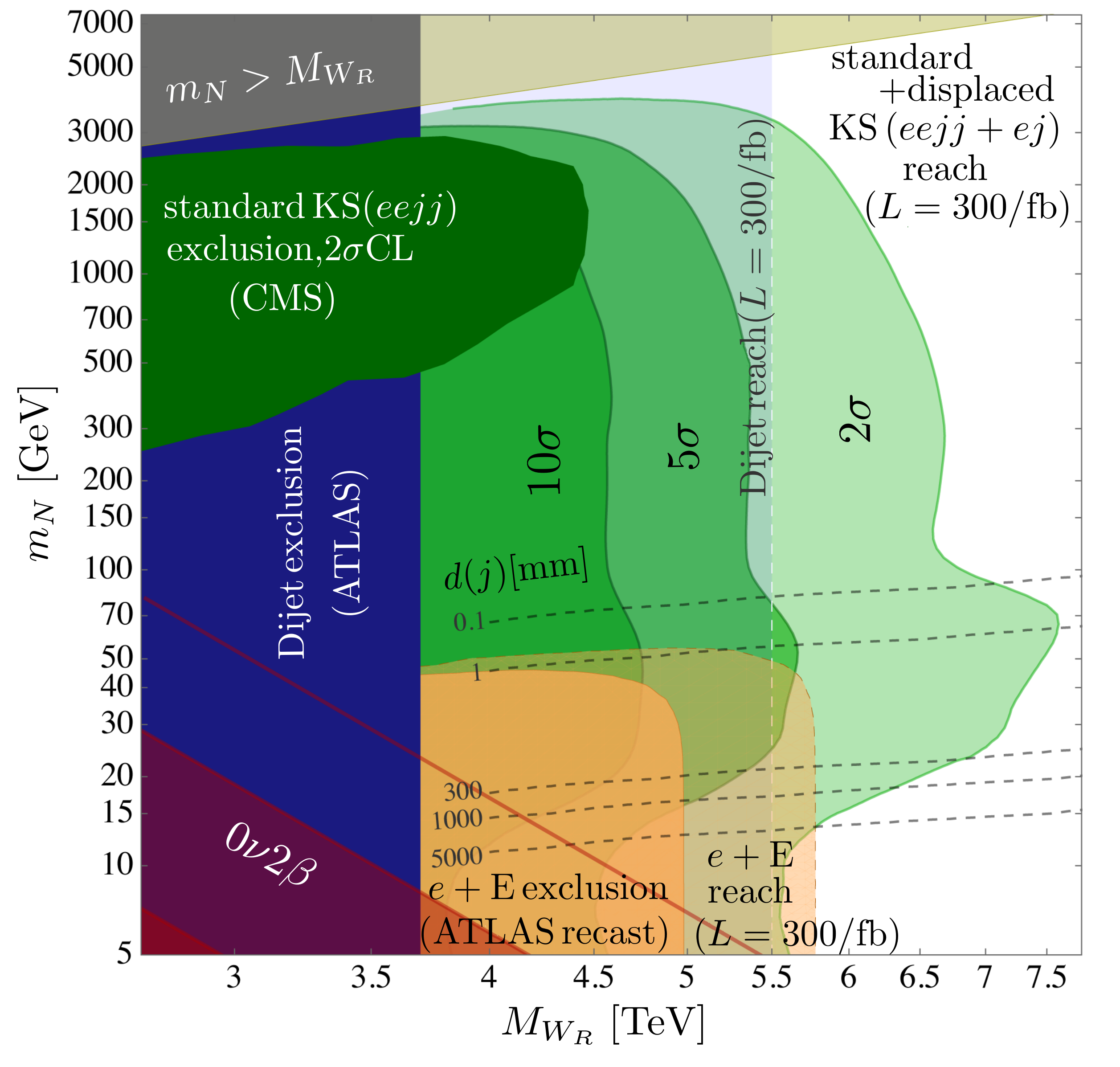}  }
\caption{Summary of LHC reach from~\cite{Nemevsek:2018bbt}.}
\label{fig-nnp}
\end{figure}

The KS process puts a limit on the $W_R$ mass of $4-5$ TeV~\cite{Aaboud:2019wfg}, depending on the mass of the RH neutrino. 
At the same time, there is a hard limit on $W_R$ mass from its decay into  di-jets, of 4 TeV~\cite{Aad:2019hjw}. The limits set by the LHC are already far above those set theoretically from low energy $K - \bar K$ mixing. Fig.\ref{fig-nnp} shows a summary  the limits found in~\cite{Nemevsek:2018bbt}. It can be seen that a $W_R$ signal could be seen all the way up to 8 TeV, with possibility of discovery up to 6 TeV. 

A similar study has been performed for a future 100 TeV collider~\cite{Ruiz:2017nip}, giving the possibility of $W_R$ discovery up to 40TeV, depending on the properties of $N$. 
 
\subsection{LR symmetry: neutrino mass vs strong CP } 

Before closing, a few words about a surprising connection between the seesaw mechanism and  strong CP violation in this theory. It stems from an important constraint that parity symmetry places on quark Yukawa couplings, otherwise arbitrary in the SM - they must be hermitian. Since parity gets broken spontaneously, quark mass matrices are not necessarily hermitian. Still, the RH quark mixing matrix gets predicted~\cite{Senjanovic:2014pva} from its LH counterpart, the CKM matrix. This is a fundamental result, for it allows to make predictions for the physics of the RH gauge boson, including the KS process that plays a crucial role in neutrino physics. \\

  Moreover, this correlates strong CP violation to the neutrino sector. The point is that the consistency of the theory, based on spontaneous breaking of parity, requires the vanishing of the QCD strong CP parameter $\theta$~\cite{Mohapatra:1978fy}. This in turn makes the electro-weak contribution to strong CP - the $ \arg \det M$ term - physical~\cite{Maiezza:2014ala}, unlike the situation in the SM. In particular, it has a profound impact on the spontaneously generated CP phase, forcing it to vanish to a great precision at the tree level. It can be shown that the loop effects involve the RH neutrino $N$, providing a beautiful interplay between strong  and the leptonic CP violation~\cite{Kuchimanchi:2014ota}. The bottom line is the same condition for the smallness of LFV then keeps the strong CP parameter $\bar \theta$ under control~\cite{Senjanovic:2020int}. Strong CP violation, far from being a problem, serves an important role of reducing the parameter space of the theory, and it does it a controlled, technically natural manner.\\

 \section{Outlook} \label{outlook}
        
            The reader probably knows that there are uncountable models of neutrino mass, and the number is growing as you read this.  It is impossible to keep track of them and for a novice it may be frustrating and confusing to follow the field of model building. If you find yourself in this category, do not worry. The main message  that I tried to convey here is that among all these models, there are very few true theories, not tailor ordered {\it a posteriori}. I have presented a series of arguments that in this sense the Left-Right Symmetric Model is unique. It is analogous to the SM in being completely self-contained and predictive.\\
          
          The crucial role in all of this is played by the neutrinoless double beta decay, because it is the first step towards a probe of  a Majorana nature of neutrino mass. As we have seen, depending on the polarization of the electron in the process, it may mean that LNV at hadron colliders ends up being equally important. If the electron comes out RH, then the neutrinoless double beta decay is due to new physics beyond neutrino mass, and this new physics would be reachable at the next collider, if not at the LHC. This would give us a unique chance to probe the theory behind the Majorana nature of the neutrino. So even if you do not care about the search of a fundamental theory, you should appreciate the importance of experiments that probe LNV. They touch into the core of the mass issue. Needless to say, it is also essential to keep determining neutrino mixings and the mass scale - even if this by itself may not be so fundamental - in order to be able to test the theories at hand.\\
          
          I also tried to convince you that deep down one has  to face the question of Left-Right symmetry, since parity is at the centre of the Standard Model. It was parity breakdown that led to the V-A theory, a precursor of the SM. We could ask then, paraphrasing Weinberg~\cite{Weinberg:2009zz}: will V+A be the key?

 \subsection*{Acknowledgments}

    I am grateful to the organisers of the Neutrino 2020 conference for giving me an opportunity to offer my vision of the theory status of our field. The online nature of the conference and the unprecedentedly huge number of participants made this truly exciting and enjoyable. Thanks are due to Umberto Cotti, Gia Dvali, Alejandra Melfo and Vladimir Tello for their encouragement, 
  and help in improving the quality and clarity of my presentation. Alejandra Melfo played an important role in shaping up this manuscript and deserves special thanks.
   
    %%%%%%%%%%
\noindent

\vskip 2 cm

\def\arxiv#1[#2]{\href{http://arxiv.org/abs/#1}{[#2]}}


\begin{thebibliography}{99}


%\cite{Fritzsch:1974nn}
\bibitem{Fritzsch:1974nn}
H.~Fritzsch and P.~Minkowski,
``Unified Interactions of Leptons and Hadrons,''
Annals Phys. \textbf{93} (1975), 193-266
doi:10.1016/0003-4916(75)90211-0;
%1844 citations counted in INSPIRE as of 22 Jul 2020

%\cite{Georgi:1974my}
%\bibitem{Georgi:1974my}
H.~Georgi,
``The State of the Art --Gauge Theories,''
AIP Conf. Proc. \textbf{23} (1975), 575-582
doi:10.1063/1.2947450.
%287 citations counted in INSPIRE as of 22 Jul 2020


  %\cite{Minkowski:1977sc}
\bibitem{Minkowski:1977sc}
P.~Minkowski,
``Mu $\to$ E Gamma At A Rate Of One Out Of 1-Billion Muon Decays?,''
Phys.\ Lett.\ B {\bf 67} (1977) 421;
%%CITATION = PHLTA,B67,421;%%


%\cite{Mohapatra:1979ia}
%\bibitem{Mohapatra:1979ia}
  R.~N.~Mohapatra and G.~Senjanovi\'c,
  ``Neutrino Mass and Spontaneous Parity Violation,''
  Phys.\ Rev.\ Lett.\  {\bf 44} (1980) 912;
  %%CITATION = PRLTA,44,912;%%
  %3755 citations counted in INSPIRE as of 05 sept. 2015
  
 
%\cite{seesaw} 
%\bibitem{seesaw} 
%\cite{Glashow:1979nm}
%\bibitem{Glashow:1979nm} 
  S.~L.~Glashow,
  ``The Future of Elementary Particle Physics,''
  NATO Sci.\ Ser.\ B {\bf 61}, 687 (1980);
  %319 citations counted in INSPIRE as of 21 Aug 2015
  
  
%\cite{GellMann:1980vs}
%\bibitem{GellMann:1980vs}
  M.~Gell-Mann, P.~Ramond and R.~Slansky,
  ``Complex Spinors and Unified Theories,''
  Conf.\ Proc.\ C {\bf 790927} (1979) 315
  \arxiv{1306.4669}[arXiv:1306.4669 [hep-th]];
  %%CITATION = ARXIV:1306.4669;%%


  %\cite{Yanagida:1979as}
%\bibitem{Yanagida:1979as} 
  T.~Yanagida,
  ``Horizontal Symmetry And Masses Of Neutrinos,''
  Conf.\ Proc.\ C {\bf 7902131}, 95 (1979).
  %%CITATION = CONFP,C7902131,95;%%
  %1057 citations counted in INSPIRE as of 21 Aug 2015
  
  
%\cite{Rizzo:1981jr}
\bibitem{Rizzo:1981jr}
T.~G.~Rizzo and G.~Senjanovi\'c,
``Grand Unification and Parity Restoration at Low-energies. 2. Unification Constraints,''
Phys. Rev. D \textbf{25} (1982), 235
doi:10.1103/PhysRevD.25.235.
%99 citations counted in INSPIRE as of 27 Sep 2020

For an update after LEP, see e.g.
%\cite{Deshpande:1992au}
%\bibitem{Deshpande:1992au}
N.~G.~Deshpande, E.~Keith and P.~B.~Pal,
``Implications of LEP results for SO(10) grand unification,''
Phys. Rev. D \textbf{46} (1993), 2261-2264
doi:10.1103/PhysRevD.46.2261.
%129 citations counted in INSPIRE as of 27 Sep 2020


%\cite{Dvali:2007hz}
\bibitem{Dvali:2007hz}
G.~Dvali,
``Black Holes and Large N Species Solution to the Hierarchy Problem,''
Fortsch. Phys. \textbf{58}, 528-536 (2010)
doi:10.1002/prop.201000009
  \arxiv{0706.2050}[arXiv:0706.2050 [hep-th]]; 

%\cite{Dvali:2007wp}
%\bibitem{Dvali:2007wp}
G.~Dvali and M.~Redi,
``Black Hole Bound on the Number of Species and Quantum Gravity at LHC,''
Phys. Rev. D \textbf{77}, 045027 (2008)
doi:10.1103/PhysRevD.77.045027
\arxiv{0710.4344}[arXiv:0710.4344 [hep-th]].

For perturbative arguments, see e.g.
%\cite{Dvali:2001gx}
%\bibitem{Dvali:2001gx}
G.~R.~Dvali, G.~Gabadadze, M.~Kolanovi\'c and F.~Nitti,
``Scales of gravity,''
Phys. Rev. D \textbf{65} (2002), 024031
doi:10.1103/PhysRevD.65.024031
\arxiv{0106058}[arXiv:hep-th/0106058 [hep-th]];
%161 citations counted in INSPIRE as of 04 Aug 2020

%\cite{Veneziano:2001ah}
%\bibitem{Veneziano:2001ah}
G.~Veneziano,
``Large N bounds on, and compositeness limit of, gauge and gravitational interactions,''
JHEP \textbf{06}, 051 (2002)
doi:10.1088/1126-6708/2002/06/051
\arxiv{0110129}[arXiv:hep-th/0110129 [hep-th]]; 

%\cite{Han:2004wt}
%\bibitem{Han:2004wt}
T.~Han and S.~Willenbrock,
``Scale of quantum gravity,''
Phys. Lett. B \textbf{616} (2005), 215-220
doi:10.1016/j.physletb.2005.04.040
\arxiv{0404182}[arXiv:hep-ph/0404182 [hep-ph]].
%63 citations counted in INSPIRE as of 20 Jul 2020



%\cite{gg}
\bibitem{gg}
G.~Dvali and G.~Senjanovi\'c, to appear.

 %\cite{Dvali:2020wqi}
\bibitem{Dvali:2020wqi}
G.~Dvali,
``Entropy Bound and Unitarity of Scattering Amplitudes,''
\arxiv{2003.05546}[arXiv:2003.05546 [hep-th]].
%5 citations counted in INSPIRE as of 04 Aug 2020




%1253 citations counted in INSPIRE as of 27 Au
 %\cite{Dimopoulos:1981yj}
\bibitem{Dimopoulos:1981yj}
S.~Dimopoulos, S.~Raby and F.~Wilczek,
``Supersymmetry and the Scale of Unification,''
Phys. Rev. D \textbf{24} (1981), 1681-1683
doi:10.1103/PhysRevD.24.1681.
%1255 citations counted in INSPIRE as of 15 Sep 2020

\bibitem{Marciano:1981un}
W.~J.~Marciano and G.~Senjanovi\'c,
``Predictions of Supersymmetric Grand Unified Theories,''
Phys. Rev. D \textbf{25} (1982), 3092
doi:10.1103/PhysRevD.25.3092.
%417 citations counted in INSPIRE as of 04 Aug 2020


%\cite{Aulakh:1999cd}
\bibitem{Aulakh:1999cd}
C.~S.~Aulakh, A.~Melfo, A.~Ra\v{s}in and G.~Senjanovi\'c,
``Seesaw and supersymmetry or exact R-parity,''
Phys. Lett. B \textbf{459} (1999), 557-562
doi:10.1016/S0370-2693(99)00708-X
\arxiv{9902409}[arXiv:hep-ph/9902409 [hep-ph]].
%119 citations counted in INSPIRE as of 15 Sep 2020

%\cite{Aulakh:2000sn}
\bibitem{Aulakh:2000sn}
C.~S.~Aulakh, B.~Bajc, A.~Melfo, A.~Ra\v{s}in and G.~Senjanovi\'c, 
``SO(10) theory of R-parity and neutrino mass,''
Nucl. Phys. B \textbf{597} (2001), 89-109
doi:10.1016/S0550-3213(00)00721-5
\arxiv{0004031}[arXiv:hep-ph/0004031 [hep-ph]].
%169 citations counted in INSPIRE as of 15 Sep 2020


%\cite{Aulakh:1982yn}
\bibitem{Aulakh:1982yn}
C.~S.~Aulakh and R.~N.~Mohapatra,
``Neutrino as the Supersymmetric Partner of the Majoron,''
Phys. Lett. B \textbf{119} (1982), 136-140
doi:10.1016/0370-2693(82)90262-3
%520 citations counted in INSPIRE as of 15 Sep 2020


%\cite{Aulakh:1998nn}
\bibitem{Aulakh:1998nn}
C.~S.~Aulakh, A.~Melfo and G.~Senjanovi\'c,
``Minimal supersymmetric left-right model,''
Phys. Rev. D \textbf{57} (1998), 4174-4178
doi:10.1103/PhysRevD.57.4174
\arxiv{9707256}[arXiv:hep-ph/9707256 [hep-ph]];
%217 citations counted in INSPIRE as of 26 Sep 2020




%  
 %
\bibitem{Pati:1974yy} 
  J.~C.~Pati and A.~Salam,
  ``Lepton Number as the Fourth Color,''
  Phys.\ Rev.\ D {\bf 10}, 275 (1974)
  Erratum: [Phys.\ Rev.\ D {\bf 11}, 703 (1975)].
  doi:10.1103/PhysRevD.10.275, 10.1103/PhysRevD.11.703.2;

%\cite{Mohapatra:1974gc}
%\bibitem{Mohapatra:1974gc} 
  R.~N.~Mohapatra and J.~C.~Pati,
  ``A Natural Left-Right Symmetry,''
  Phys.\ Rev.\ D {\bf 11}, 2558 (1975).
  doi:10.1103/PhysRevD.11.2558;
  %%CITATION = doi:10.1103/PhysRevD.11.2558;%%
 
 %\cite{Senjanovic:1975rk}
%\bibitem{Senjanovic:1975rk}
G.~Senjanovi\'c and R.~N.~Mohapatra,
``Exact Left-Right Symmetry And Spontaneous Violation Of Parity,''
Phys.\ Rev.\ D {\bf 12} (1975) 1502;
%%CITATION = PHRVA,D12,1502;%%


%\cite{Senjanovic:1978ev}
%\bibitem{Senjanovic:1978ev} 
  G.~Senjanovi\'c,
  ``Spontaneous Breakdown of Parity in a Class of Gauge Theories,''
  Nucl.\ Phys.\ B {\bf 153}, 334 (1979).
  doi:10.1016/0550-3213(79)90604-7.
  %%CITATION = doi:10.1016/0550-3213(79)90604-7;%%
  
  
%\cite{Slansky:1981yr}
\bibitem{Slansky:1981yr}
R.~Slansky,
``Group Theory for Unified Model Building,''
Phys. Rept. \textbf{79} (1981), 1-128
doi:10.1016/0370-1573(81)90092-2;
%1240 citations counted in INSPIRE as of 10 Oct 2020

%\cite{Kibble:1982ae}
%\bibitem{Kibble:1982ae}
T.~W.~B.~Kibble, G.~Lazarides and Q.~Shafi,
``Strings in SO(10),''
Phys. Lett. B \textbf{113} (1982), 237-239
doi:10.1016/0370-2693(82)90829-2
%195 citations counted in INSPIRE as of 10 Oct 2020


  
%\cite{Akhmedov:1992hh}
\bibitem{Akhmedov:1992hh}
E.~K.~Akhmedov, Z.~G.~Berezhiani and G.~Senjanovi\'c,
``Planck scale physics and neutrino masses,''
Phys. Rev. Lett. \textbf{69} (1992), 3013-3016
doi:10.1103/PhysRevLett.69.3013
\arxiv{9205230}[arXiv:hep-ph/9205230 [hep-ph]].
%205 citations counted in INSPIRE as of 04 Aug 2020

%\cite{Dvali:2013cpa}
\bibitem{Dvali:2013cpa}
G.~Dvali, S.~Folkerts and A.~Franca,
``How neutrino protects the axion,''
Phys. Rev. D \textbf{89} (2014) no.10, 105025
doi:10.1103/PhysRevD.89.105025
\arxiv{1312.7273}[arXiv:1312.7273 [hep-th]];
%32 citations counted in INSPIRE as of 04 Aug 2020

%\cite{Dvali:2016uhn}
%\bibitem{Dvali:2016uhn}
G.~Dvali and L.~Funcke,
``Small neutrino masses from gravitational $\theta$ term,''
Phys. Rev. D \textbf{93} (2016) no.11, 113002
doi:10.1103/PhysRevD.93.113002
\arxiv{1602.03191}[arXiv:1602.03191 [hep-ph]].
%55 citations counted in INSPIRE as of 04 Aug 2020

%\cite{Dvali:2017mpy}
\bibitem{Dvali:2017mpy}
G.~Dvali,
``Topological Origin of Chiral Symmetry Breaking in QCD and in Gravity,''
\arxiv{1705.06317 }[arXiv:1705.06317 [hep-th]].
%9 citations counted in INSPIRE as of 04 Aug 2020


%\cite{Majorana:1937vz}
\bibitem{Majorana:1937vz}
  E.~Majorana,
  ``Theory Of The Symmetry Of Electrons And Positrons,''
  Nuovo Cim.\  {\bf 14}, 171 (1937).
  %%CITATION = NUCIA,14,171;%%

 
%\cite{Racah:1937qq}
\bibitem{Racah:1937qq}
  G.~Racah,
  ``On the symmetry of particle and antiparticle,''
  Nuovo Cim.\  {\bf 14}, 322 (1937);
  %%CITATION = NUCIA,14,322;%%

%\cite{Furry:1939qr}
%\bibitem{Furry:1939qr}
  W.~H.~Furry,
  ``On transition probabilities in double beta-disintegration,''
  Phys.\ Rev.\  {\bf 56}, 1184 (1939).
  %%CITATION = PHRVA,56,1184;%%
  
  For a review, see e.g.
  %\cite{GomezCadenas:2011it}
%\bibitem{GomezCadenas:2011it}
J.~J.~Gomez-Cadenas, J.~Martin-Albo, M.~Mezzetto, F.~Monrabal and M.~Sorel,
``The Search for neutrinoless double beta decay,''
Riv. Nuovo Cim. \textbf{35} (2012) no.2, 29-98
doi:10.1393/ncr/i2012-10074-9
\arxiv{1109.5515}[arXiv:1109.5515 [hep-ex]].
%212 citations counted in INSPIRE as of 09 Oct 2020
  
  For a recent status, see e.g.
  %\cite{Dolinski:2019nrj}
%\bibitem{Dolinski:2019nrj}
M.~J.~Dolinski, A.~W.~P.~Poon and W.~Rodejohann,
``Neutrinoless Double-Beta Decay: Status and Prospects,''
Ann. Rev. Nucl. Part. Sci. \textbf{69} (2019), 219-251
doi:10.1146/annurev-nucl-101918-023407
\arxiv{1902.04097}[arXiv:1902.04097 [nucl-ex]].
%93 citations counted in INSPIRE as of 09 Oct 2020
  
  
%\cite{Agostini:2020xta}
\bibitem{Agostini:2020xta}
``M.~Agostini \textit{et al.} [GERDA],
Final Results of GERDA on the Search for Neutrinoless Double-$\beta$ Decay,''
\arxiv{2009.06079}[arXiv:2009.06079 [nucl-ex]].
%0 citations counted in INSPIRE as of 15 Sep 2020}\



%\cite{Schechter:1981bd}
\bibitem{Schechter:1981bd}
J.~Schechter and J.~W.~F.~Valle,
``Neutrinoless Double beta Decay in SU(2) x U(1) Theories,''
Phys. Rev. D \textbf{25} (1982), 2951
doi:10.1103/PhysRevD.25.2951
%826 citations counted in INSPIRE as of 10 Aug 2020


%\cite{Duerr:2011zd}
\bibitem{Duerr:2011zd}
M.~Duerr, M.~Lindner and A.~Merle,
``On the Quantitative Impact of the Schechter-Valle Theorem,''
JHEP \textbf{06} (2011), 091
doi:10.1007/JHEP06(2011)091
\arxiv{1105.0901}[arXiv:1105.0901 [hep-ph]].
%108 citations counted in INSPIRE as of 10 Aug 2020





%\cite{maurice}
    \bibitem{maurice}
	G.~Feinberg, M.~Goldhaber,
	Proc.\ Nat.\ Ac.\ Sci.\ USA {\bf 45} (1959) 1301;
	
%\cite{Pontecorvo:1968wp}
%\bibitem{Pontecorvo:1968wp}
  B.~Pontecorvo,
  ``Superweak interactions and double beta decay,''
 Phys.\ Lett.\  {\bf B26 } (1968)  630.

 
For the modern gauge theory point of view, see 
%\cite{Mohapatra:1980yp}
%\bibitem{Mohapatra:1980yp}
R.~N.~Mohapatra and G.~Senjanovi\'c,
``Neutrino Masses and Mixings in Gauge Models with Spontaneous Parity Violation,''
Phys. Rev. D \textbf{23} (1981), 165
doi:10.1103/PhysRevD.23.165
%2578 citations counted in INSPIRE as of 26 Sep 2020


 
 
 %\cite{Keung:1983uu}
\bibitem{Keung:1983uu}
  W.~Y.~Keung and G.~Senjanovi\'c,
  ``Majorana Neutrinos And The Production Of The Right-Handed Charged Gauge
  Boson,''
  Phys.\ Rev.\ Lett.\  {\bf 50}, 1427 (1983).
  %%CITATION = PRLTA,50,1427;%%
  
   %\cite{Aaboud:2019wfg}
\bibitem{Aaboud:2019wfg} 
  M.~Aaboud {\it et al.} [ATLAS Collaboration],
  ``Search for a right-handed gauge boson decaying into a high-momentum heavy neutrino and a charged lepton in $pp$ collisions with the ATLAS detector at $\sqrt{s}=13$ TeV,''
  Phys.\ Lett.\ B {\bf 798}, 134942 (2019)
  doi:10.1016/j.physletb.2019.134942
\arxiv{1904.12679}  [arXiv:1904.12679 [hep-ex]];
  %%CITATION = doi:10.1016/j.physletb.2019.134942;%%
  
  %\cite{Sirunyan:2018pom}
\bibitem{Sirunyan:2018pom}
A.~M.~Sirunyan \textit{et al.} [CMS],
``Search for a heavy right-handed W boson and a heavy neutrino in events with two same-flavor leptons and two jets at $\sqrt{s}=$ 13 TeV,''
JHEP \textbf{05} (2018), 148
doi:10.1007/JHEP05(2018)148
\arxiv{1803.11116}[arXiv:1803.11116 [hep-ex]].
%39 citations counted in INSPIRE as of 11 Oct 2020
  
  
  
  
  %\cite{Lee:1956qn}
\bibitem{Lee:1956qn}
T.~D.~Lee and C.~N.~Yang,
``Question of Parity Conservation in Weak Interactions,''
Phys. Rev. \textbf{104} (1956), 254-258
doi:10.1103/PhysRev.104.254
%1825 citations counted in INSPIRE as of 04 Aug 2020
  
  
  %\cite{Sudarshan:1958vf}
\bibitem{Sudarshan:1958vf}
E.~C.~G.~Sudarshan and R.~E.~Marshak,
``Chirality invariance and the universal Fermi interaction,''
Phys. Rev. \textbf{109} (1958), 1860-1860
doi:10.1103/PhysRev.109.1860.2;
%595 citations counted in INSPIRE as of 10 Aug 2020

%\cite{Feynman:1958ty}
%\bibitem{Feynman:1958ty}
R.~P.~Feynman and M.~Gell-Mann,
``Theory of Fermi interaction,''
Phys. Rev. \textbf{109} (1958), 193-198
doi:10.1103/PhysRev.109.193
%1695 citations counted in INSPIRE as of 10 Aug 2020


%\cite{Sirunyan:2020two}
\bibitem{Sirunyan:2020two}
A.~M.~Sirunyan \textit{et al.} [CMS],
``Evidence for Higgs boson decay to a pair of muons,''
[arXiv:2009.04363 [hep-ex]].
%0 citations counted in INSPIRE as of 20 Sep 2020

  
  %\cite{Weinberg:2009zz}
\bibitem{Weinberg:2009zz}
S.~Weinberg,
``V-A was the key,''
J. Phys. Conf. Ser. \textbf{196} (2009), 012002
doi:10.1088/1742-6596/196/1/012002
%18 citations counted in INSPIRE as of 05 Aug 2020


%\cite{Branco:1978bz}
\bibitem{Branco:1978bz}
G.~C.~Branco and G.~Senjanovi\'c,
``The Question of Neutrino Mass,''
Phys. Rev. D \textbf{18} (1978), 1621
doi:10.1103/PhysRevD.18.1621.
%62 citations counted in INSPIRE as of 05 Aug 2020

For attempts to revive the Dirac neutrino mass case, see e.g.
%\cite{Mishra:1987tz}
%\bibitem{Mishra:1987tz}
S.~Mishra, S.~P.~Misra, S.~Panda and U.~Sarkar,
``Light Dirac Neutrino in Left-right Symmetric Models,''
Phys. Rev. D \textbf{35} (1987), 975
doi:10.1103/PhysRevD.35.975;
%6 citations counted in INSPIRE as of 10 Oct 2020

%\cite{Bolton:2019bou}
%\bibitem{Bolton:2019bou}
P.~D.~Bolton, F.~F.~Deppisch, C.~Hati, S.~Patra and U.~Sarkar,
``Alternative formulation of left-right symmetry with $B-L$ conservation and purely Dirac neutrinos,''
Phys. Rev. D \textbf{100} (2019) no.3, 035013
doi:10.1103/PhysRevD.100.035013
\arxiv{1902.05802}[arXiv:1902.05802 [hep-ph]].
%14 citations counted in INSPIRE as of 10 Oct 2020

%\cite{Chavez:2019yal}
%\bibitem{Chavez:2019yal}
H.~Diaz Chavez, V.~Pleitez and O.~P.~Ravinez,
``Dirac neutrinos in a $SU(2)$ left-right symmetric model,''
Phys. Rev. D \textbf{102} (2020), 075006
doi:10.1103/PhysRevD.102.075006
\arxiv{1908.02828}[arXiv:1908.02828 [hep-ph]].
%4 citations counted in INSPIRE as of 12 Oct 2020

 
    %\cite{Nemevsek:2012iq}
\bibitem{Nemevsek:2012iq}
  M.~Nemev\v{s}ek, G.~Senjanovi\'c and V.~Tello,
  ``Connecting Dirac and Majorana Neutrino Mass Matrices in the Minimal Left-Right Symmetric Model,''
  Phys.\ Rev.\ Lett.\  {\bf 110} (2013) 15,  151802
 \arxiv{1902.05802}[arXiv:1211.2837 [hep-ph]];
  %%CITATION = ARXIV:1211.2837;%%

 %\cite{Senjanovic:2016vxw}
%\bibitem{Senjanovic:2016vxw}
G.~Senjanovi\'c and V.~Tello,
``Probing Seesaw with Parity Restoration,''
Phys. Rev. Lett. \textbf{119} (2017) no.20, 201803
doi:10.1103/PhysRevLett.119.201803
\arxiv{1612.05503}[arXiv:1612.05503 [hep-ph]].
%13 citations counted in INSPIRE as of 07 Aug 2020

%\cite{Casas:2001sr}
\bibitem{Casas:2001sr}
J.~A.~Casas and A.~Ibarra,
``Oscillating neutrinos and $\mu \to e, \gamma$,''
Nucl. Phys. B \textbf{618} (2001), 171-204
doi:10.1016/S0550-3213(01)00475-8
\arxiv{0103065}[arXiv:hep-ph/0103065 [hep-ph]].
%1074 citations counted in INSPIRE as of 08 Oct 2020

     
%\cite{Senjanovic:2011zz}
\bibitem{Senjanovic:2011zz}
G.~Senjanovi\'c,
``Neutrino mass: From LHC to grand unification,''
Riv. Nuovo Cim. \textbf{34} (2011) no.1, 1-68
doi:10.1393/ncr/i2011-10061-8
%45 citations counted in INSPIRE as of 03 Aug 2020

%\cite{Gluza:2016qqv}
\bibitem{Gluza:2016qqv}
For a detailed study, see e.g. J.~Gluza, T.~Jelinski and R.~Szafron,
``Lepton number violation and \textquoteleft{}Diracness\textquoteright{} of massive neutrinos composed of Majorana states,''
Phys. Rev. D \textbf{93} (2016) no.11, 113017
doi:10.1103/PhysRevD.93.113017
\arxiv{1604.01388}[arXiv:1604.01388 [hep-ph]].
%54 citations counted in INSPIRE as of 09 Oct 2020

  
%\cite{Tello:2010am}
\bibitem{Tello:2010am}
V.~Tello, M.~Nemev\v{s}ek, F.~Nesti, G.~Senjanovi\'c and F.~Vissani,
``Left-Right Symmetry: from LHC to Neutrinoless Double Beta Decay,''
Phys. Rev. Lett. \textbf{106} (2011), 151801
doi:10.1103/PhysRevLett.106.151801
\arxiv{1011.3522}[arXiv:1011.3522 [hep-ph]];
%220 citations counted in INSPIRE as of 08 Aug 2020

%\cite{Nemevsek:2011aa}
%\bibitem{Nemevsek:2011aa}
M.~Nemev\v{s}ek, F.~Nesti, G.~Senjanovi\'c and V.~Tello,
``Neutrinoless Double Beta Decay: Low Left-Right Symmetry Scale?,''
\arxiv{1112.3061}[arXiv:1112.3061 [hep-ph]].
%70 citations counted in INSPIRE as of 27 Aug 2020

See also,
%\cite{Deppisch:2012nb}
%\bibitem{Deppisch:2012nb}
F.~F.~Deppisch, M.~Hirsch and H.~Pas,
``Neutrinoless Double Beta Decay and Physics Beyond the Standard Model,''
J. Phys. G \textbf{39} (2012), 124007
doi:10.1088/0954-3899/39/12/124007
\arxiv{1208.0727}[arXiv:1208.0727 [hep-ph]];
%189 citations counted in INSPIRE as of 08 Oct 2020

%\cite{Chakrabortty:2012mh}
%\bibitem{Chakrabortty:2012mh}
J.~Chakrabortty, H.~Z.~Devi, S.~Goswami and S.~Patra,
``Neutrinoless double-$\beta$ decay in TeV scale Left-Right symmetric models,''
JHEP \textbf{08} (2012), 008
doi:10.1007/JHEP08(2012)008
\arxiv{1204.2527}[arXiv:1204.2527 [hep-ph]];
%76 citations counted in INSPIRE as of 08 Oct 2020

%\cite{Huang:2013kma}
%\bibitem{Huang:2013kma}
W.~C.~Huang and J.~Lopez-Pavon,
``On neutrinoless double beta decay in the minimal left-right symmetric model,''
Eur. Phys. J. C \textbf{74} (2014), 2853
doi:10.1140/epjc/s10052-014-2853-z
\arxiv{1310.0265}[arXiv:1310.0265 [hep-ph]].
%36 citations counted in INSPIRE as of 08 Oct 2020


%\cite{Li:2020flq}
\bibitem{Li:2020flq}
G.~Li, M.~Ramsey-Musolf and J.~C.~Vasquez,
``Left-right symmetry and leading contributions to neutrinoless double beta decay,''
\arxiv{2009.01257}[arXiv:2009.01257 [hep-ph]].
%0 citations counted in INSPIRE as of 09 Oct 2020

%\cite{Cirigliano:2004mv}
\bibitem{Cirigliano:2004mv}
V.~Cirigliano, A.~Kurylov, M.~J.~Ramsey-Musolf and P.~Vogel,
``Lepton flavor violation without supersymmetry,''
Phys. Rev. D \textbf{70} (2004), 075007
doi:10.1103/PhysRevD.70.075007
\arxiv{0404233}[arXiv:hep-ph/0404233 [hep-ph]].
%87 citations counted in INSPIRE as of 09 Oct 2020

 
  %\cite{Tello:2012qda}
\bibitem{Tello:2012qda}
V.~Tello,
``Connections between the high and low energy violation of Lepton and Flavor numbers in the minimal left-right symmetric model,''
%3 citations counted in INSPIRE as of 08 Aug 2020

%\cite{Datta:1993nm}
\bibitem{Datta:1993nm}
A.~Datta, M.~Guchait and A.~Pilaftsis,
``Probing lepton number violation via majorana neutrinos at hadron supercolliders,''
Phys. Rev. D \textbf{50} (1994), 3195-3203
doi:10.1103/PhysRevD.50.3195
\arxiv{1902.05802}[arXiv:hep-ph/9311257 [hep-ph]].
%166 citations counted in INSPIRE as of 08 Oct 2020
  
 
%\cite{Senjanovic:2018xtu}
\bibitem{Senjanovic:2018xtu}
G.~Senjanovi\'c and V.~Tello,
``Disentangling the seesaw mechanism in the minimal left-right symmetric model,''
Phys. Rev. D \textbf{100} (2019) no.11, 115031
doi:10.1103/PhysRevD.100.115031
\arxiv{1812.03790}[arXiv:1812.03790 [hep-ph]];
%5 citations counted in INSPIRE as of 07 Aug 2020

%\cite{Senjanovic:2019moe}
%\bibitem{Senjanovic:2019moe}
G.~Senjanovi\'c and V.~Tello,
``Parity and the origin of neutrino mass,''
Int. J. Mod. Phys. A \textbf{35} (2020) no.09, 2050053
doi:10.1142/S0217751X20500530
\arxiv{1912.13060}[arXiv:1912.13060 [hep-ph]].
%1 citations counted in INSPIRE as of 08 Aug 2020

%\cite{Arbelaez:2017zqq}
\bibitem{Arbelaez:2017zqq}
C.~Arbel\'aez, C.~Dib, I.~Schmidt and J.~C.~Vasquez,
``Probing the Dirac or Majorana nature of the Heavy Neutrinos in pure leptonic decays at the LHC,''
Phys. Rev. D \textbf{97} (2018) no.5, 055011
doi:10.1103/PhysRevD.97.055011
\arxiv{1712.08704}[arXiv:1712.08704 [hep-ph]];
%9 citations counted in INSPIRE as of 27 Aug 2020

%\cite{Helo:2018rll}
%\bibitem{Helo:2018rll}
J.~C.~Helo, H.~Li, N.~A.~Neill, M.~Ramsey-Musolf and J.~C.~Vasquez,
``Probing neutrino Dirac mass in left-right symmetric models at the LHC and next generation colliders,''
Phys. Rev. D \textbf{99} (2019) no.5, 055042
doi:10.1103/PhysRevD.99.055042
\arxiv{1812.01630}[arXiv:1812.01630 [hep-ph]],
%8 citations counted in INSPIRE as of 09 Oct 2020

 

%\cite{Ferrari:2000sp}
\bibitem{Ferrari:2000sp}
A.~Ferrari, J.~Collot, M.~L.~Andrieux, B.~Belhorma, P.~de Saintignon, J.~Y.~Hostachy, P.~Martin and M.~Wielers,
``Sensitivity study for new gauge bosons and right-handed Majorana neutrinos in $p p$ collisions at $s$ = 14-TeV,''
Phys. Rev. D \textbf{62} (2000), 013001
doi:10.1103/PhysRevD.62.013001;
%155 citations counted in INSPIRE as of 27 Aug 2020

%\cite{Han:2012vk}
%\bibitem{Han:2012vk}
T.~Han, I.~Lewis, R.~Ruiz and Z.~g.~Si,
``Lepton Number Violation and $W^\prime$ Chiral Couplings at the LHC,''
Phys. Rev. D \textbf{87} (2013) no.3, 035011
doi:10.1103/PhysRevD.87.035011
\arxiv{1211.6447}[arXiv:1211.6447 [hep-ph]].
%68 citations counted in INSPIRE as of 27 Aug 2020

%\cite{Aad:2019hjw}
\bibitem{Aad:2019hjw}
G.~Aad \textit{et al.} [ATLAS],
``Search for new resonances in mass distributions of jet pairs using 139 fb$^{-1}$ of $pp$ collisions at $\sqrt{s}=13$ TeV with the ATLAS detector,''
JHEP \textbf{03} (2020), 145
doi:10.1007/JHEP03(2020)145
\arxiv{1910.08447}[arXiv:1910.08447 [hep-ex]].
%38 citations counted in INSPIRE as of 11 Oct 2020


  
  %\cite{Nemevsek:2018bbt}
\bibitem{Nemevsek:2018bbt} 
  M.~Nemev\v{s}ek, F.~Nesti and G.~Popara,
  ``Keung-Senjanovi\'c process at the LHC: From lepton number violation to displaced vertices to invisible decays,''
  Phys.\ Rev.\ D {\bf 97}, no. 11, 115018 (2018)
  doi:10.1103/PhysRevD.97.115018
 \arxiv{1801.05813}[arXiv:1801.05813 [hep-ph]].
  %%CITATION = doi:10.1103/PhysRevD.97.115018;%%



  
 
%\cite{Ruiz:2017nip}
\bibitem{Ruiz:2017nip}
R.~Ruiz,
``Lepton Number Violation at Colliders from Kinematically Inaccessible Gauge Bosons,''
Eur. Phys. J. C \textbf{77} (2017) no.6, 375
doi:10.1140/epjc/s10052-017-4950-2
\arxiv{1703.04669}[arXiv:1703.04669 [hep-ph]].
%25 citations counted in INSPIRE as of 07 Aug 2020

%\cite{Senjanovic:2014pva}
\bibitem{Senjanovic:2014pva}
G.~Senjanovi\'c and V.~Tello,
``Right Handed Quark Mixing in Left-Right Symmetric Theory,''
Phys. Rev. Lett. \textbf{114} (2015) no.7, 071801
doi:10.1103/PhysRevLett.114.071801
\arxiv{1408.3835}[arXiv:1408.3835 [hep-ph]];
%59 citations counted in INSPIRE as of 08 Aug 2020

%\cite{Senjanovic:2015yea}
%\bibitem{Senjanovic:2015yea}
G.~Senjanovi\'c and V.~Tello,
``Restoration of Parity and the Right-Handed Analog of the CKM Matrix,''
Phys. Rev. D \textbf{94} (2016) no.9, 095023
doi:10.1103/PhysRevD.94.095023
\arxiv{1502.05704}[arXiv:1502.05704 [hep-ph]].
%39 citations counted in INSPIRE as of 07 Aug 2020

For an illustrative application, see
%\cite{Bertolini:2019out}
%\bibitem{Bertolini:2019out}
S.~Bertolini, A.~Maiezza and F.~Nesti,
``Kaon CP violation and neutron EDM in the minimal left-right symmetric model,''
Phys. Rev. D \textbf{101} (2020) no.3, 035036
doi:10.1103/PhysRevD.101.035036
\arxiv{1911.09472}[arXiv:1911.09472 [hep-ph]].
%8 citations counted in INSPIRE as of 27 Aug 2020

For earlier work, see
%\cite{Zhang:2007fn}
%\bibitem{Zhang:2007fn}
Y.~Zhang, H.~An, X.~Ji and R.~N.~Mohapatra,
``Right-handed quark mixings in minimal left-right symmetric model with general CP violation,''
Phys. Rev. D \textbf{76} (2007), 091301
doi:10.1103/PhysRevD.76.091301
\arxiv{0704.1662}[arXiv:0704.1662 [hep-ph]].
%71 citations counted in INSPIRE as of 27 Aug 2020

%\cite{Mohapatra:1978fy}
\bibitem{Mohapatra:1978fy}
R.~N.~Mohapatra and G.~Senjanovi\'c,
``Natural Suppression of Strong p and t Noninvariance,''
Phys. Lett. B \textbf{79} (1978), 283-286
doi:10.1016/0370-2693(78)90243-5
%234 citations counted in INSPIRE as of 27 Sep 2020

%\cite{Maiezza:2014ala}
\bibitem{Maiezza:2014ala}
A.~Maiezza and M.~Nemev\v{s}ek,
``Strong P invariance, neutron electric dipole moment, and minimal left-right parity at LHC,''
Phys. Rev. D \textbf{90} (2014) no.9, 095002
doi:10.1103/PhysRevD.90.095002
\arxiv{1407.3678}[arXiv:1407.3678 [hep-ph]].
%59 citations counted in INSPIRE as of 27 Sep 2020

%\cite{Kuchimanchi:2014ota}
\bibitem{Kuchimanchi:2014ota}
R.~Kuchimanchi,
``Leptonic CP problem in left-right symmetric model,''
Phys. Rev. D \textbf{91} (2015) no.7, 071901
doi:10.1103/PhysRevD.91.071901
\arxiv{1408.6382}[arXiv:1408.6382 [hep-ph]].
%14 citations counted in INSPIRE as of 27 Sep 2020

%\cite{Senjanovic:2020int}
\bibitem{Senjanovic:2020int}
G.~Senjanovi\'c and V.~Tello,
``Strong CP violation: problem or blessing?,''
\arxiv{2004.04036}[arXiv:2004.04036 [hep-ph]].
%2 citations counted in INSPIRE as of 27 Sep 2020





\end{thebibliography}
\end{document}